\documentclass[a4paper,11pt]{article}
\usepackage{pos}

\title{New electroweak challenges and opportunities at the LHeC}

\author*[a]{K. Piotrzkowski}
\author[b]{Y. Yamazaki}

\affiliation[a]{Centre for Cosmology, Particle Physics and Phenomenology (CP$^3$),
Université catholique de Louvain
\\
  Chemin du Cyclotron 2,
B-1348 Louvain-la-Neuve,
Belgium}

\affiliation[b]{Physics Department, Graduate School of Science, Kobe University,\\1-1 Rokkodai-cho, Nada-ku, Kobe, Japan}

\emailAdd{krzysztof.piotrzkowski@cern.ch}
\emailAdd{yamazaki@phys.sci.kobe-u.ac.jp}

\abstract{The Large Hadron-Electron Collider (LHeC) will operate at $\sqrt{s}$ = 1.2 TeV and accumulate about 1/ab of integrated electron-proton luminosity. Novel studies of high energy photon-photon interactions at the LHeC, at the $\gamma\gamma$ center-of-mass energy of up to 1 TeV, will open new frontiers in the electroweak physics as well as in searches for physics beyond the Standard Model. Despite a very high $ep$ luminosity, the experimental conditions will be very favorable at the LHeC – a negligible event pileup will allow for unique studies of a number of processes involving the exclusive production via two-photon fusion.}

\FullConference{%
  *** The European Physical Society Conference on High Energy Physics (EPS-HEP2021), ***\\
  *** 26-30 July 2021 ***\\
  *** Online conference, jointly organized by Universität Hamburg and the research center DESY ***
}

\begin{document}
\maketitle

\section{Introduction}
A significant fraction of all proton-proton interactions at the LHC proceeds via photon-photon fusion where the incoming protons are elastically scattered at very small angles. Initially, a practical conversion of the LHC into a high energy photon-photon collider was proposed by means of detection of such forward-scattered protons \cite{Piotrzkowski:2000rx}. It was found that the effective photon-photon luminosity reaches almost 1\% of the $pp$ luminosity for the $\gamma\gamma$ center-of-mass energy $W$ above 100~GeV, at $\sqrt{s}$ = 14~TeV, opening a new field of research, in the electroweak sector in particular.

\begin{figure}[htbp]
\includegraphics[width=1.0\textwidth]{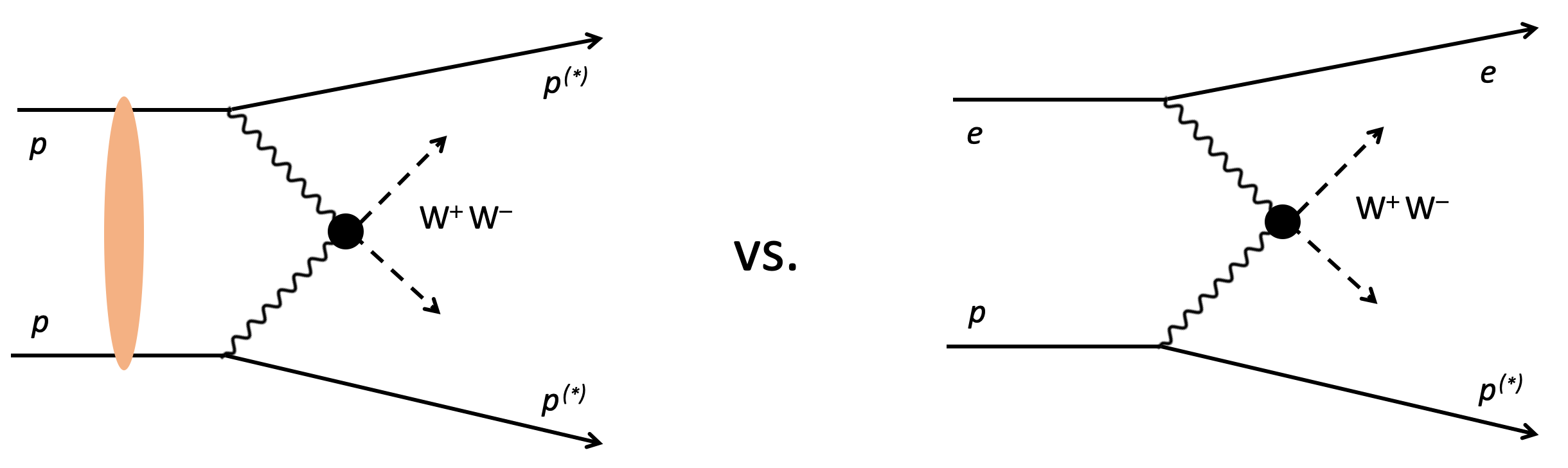}
\caption{\label{fig_1} Diagrams for the exclusive W-pair production via photon-photon fusion in proton-proton and electron-proton collisions. The symbol (*) means that the forward final state could be an intact proton, or a dissociative system due to the proton breakup. The orange blob, for the $pp$ case, represents hadron-hadron re-scattering effects.}
\end{figure}

Of course, in general, the $\gamma\gamma$ cross-sections are much smaller than the partonic ones, and a clean selection of photon-photon interactions has been highly not trivial. However, resolutions of the central trackers are so high at the LHC that such $\gamma\gamma$ interactions can be identified using a very powerful exclusivity condition -- for example, by demanding that exactly two tracks are registered at a given event vertex. Such an exclusivity selection based on tracks only has been essential here, as it could be successfully applied in the case of high event pileup conditions at the LHC. A proof-of-concept demonstration was performed by measuring the exclusive two-photon production of muon pairs, using all the available data at the time \cite{CMS:2011vma}. It was done without detecting the forward scattered protons, which permitted the additional, non-elastic contributions due to proton dissociative events (see Fig. \ref{fig_1}). The tracks in exclusivity selection were limited to the central region, $|\eta_{track}|<2.4$, which therefore allowed very high dissociative masses and resulted in an increase of the effective $\gamma\gamma$ luminosity by a factor of about 4.

This technique was used to produce at $\sqrt{s}$ = 7~TeV first ever evidence for the two-photon W-pair production, and to put unique limits on the anomalous quartic $\gamma\gamma WW$ couplings in particular \cite{CMS:2013hdf}. Recently, this has been successfully performed using the data collected at $\sqrt{s}$ = 13~TeV, and at much higher event pileup \cite{ATLAS:2020iwi}. This was done, however, still using only one very special decay mode: $WW\rightarrow e\mu\nu\nu$. This shows hard limits of this technique due to the hugely difficult conditions at the LHC. Even detection of the forward-scattered protons does not solve this problem, as the event statistics is much reduced then and also a significant event pileup is present in the forward proton detectors \cite{ATLAS:2020mve}. That difficulty will yet strongly increase at the HL-LHC, and can only be mitigated by the use of ultra-precise picosecond resolution timing detectors.

\section{High energy $\gamma\gamma$ interactions at the LHeC}
A revolution in accelerator science is taking place recently thanks to breakthroughs in developments of the {\it Energy Recovery Linac} (ERL) technology \cite{erl}. The LHeC proposal benefits from that by bringing the peak luminosity for electron-proton collisions above 10$^{34}$ cm$^{-2}$s$^{-1}$ \cite{LHeC:2020van}. It is expected that the accumulated $ep$ luminosity will be of about 1/ab. This and the $ep$ center-of-mass energy of 1.2~TeV will provide excellent conditions at the LHeC for groundbreaking studies of high energy photon-photon interactions.

\begin{figure}[htbp]
\includegraphics[width=0.92\textwidth]{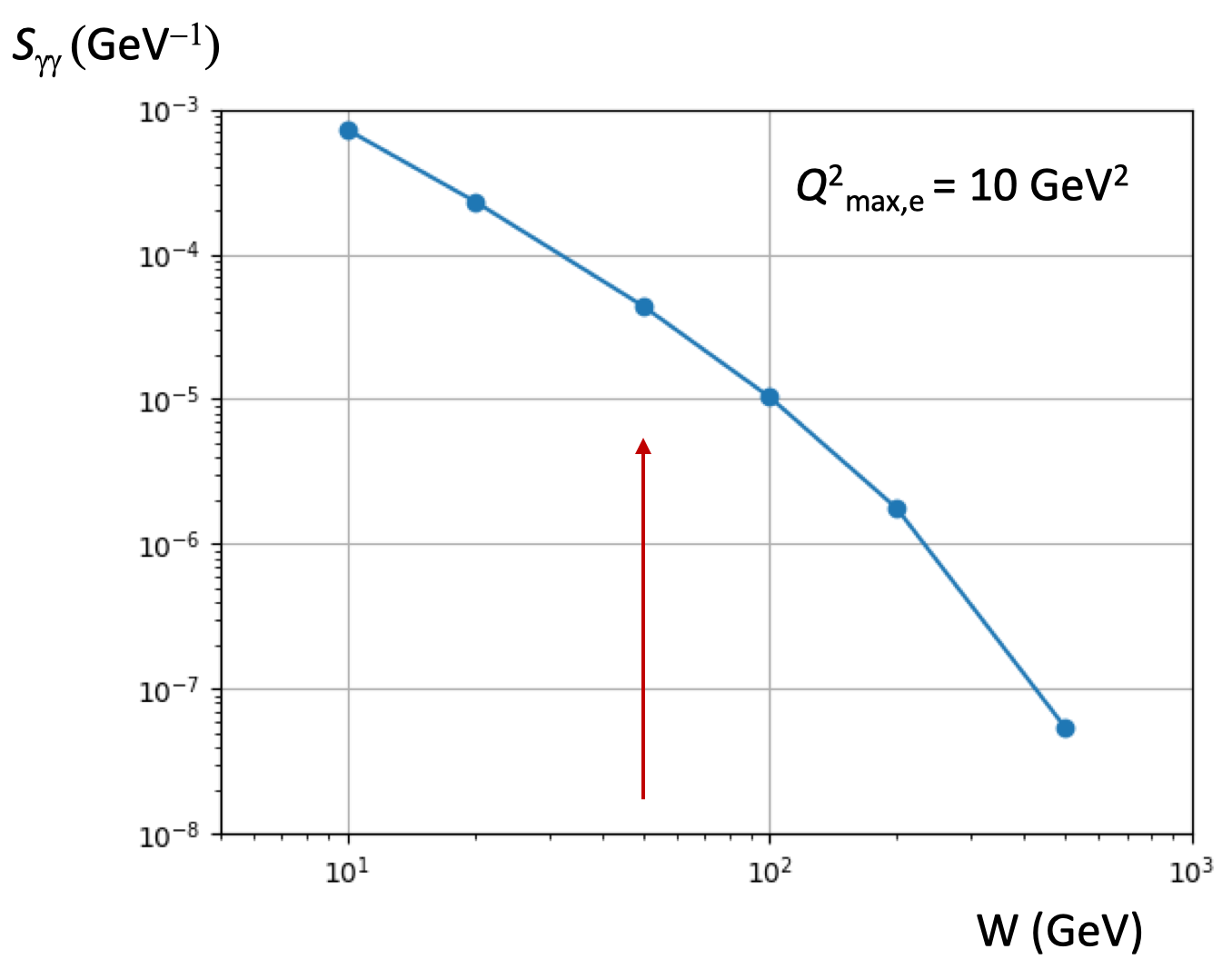}
\caption{\label{fig_2} The photon-photon luminosity spectrum $S_{\gamma\gamma}$ at the LHeC for $\sqrt{s}$ = 1.2~TeV, calculated according to the EPA \cite{Budnev:1975poe}, for the elastic events and maximum virtuality at the electron vertex of 10~GeV$^2$. The arrow indicates $S_{\gamma\gamma}$ at W = 50~GeV, where it has a similar value to the LHC one.}
\end{figure}

In the same fashion as in Ref. \cite{Piotrzkowski:2000rx}, the intensity of the effective photon-photon collisions at the LHeC can be expressed by the photon-photon luminosity spectrum $S_{\gamma\gamma}$, calculated from a convolution of two photon fluxes according the {\it Equivalent Photon Approximation} (EPA) \cite{Budnev:1975poe}. In spite of the much smaller electron beam energy of 50~GeV, the $S_{\gamma\gamma}$ in the medium $W$ range for elastic electron-proton collisions  is comparable to the elastic $S_{\gamma\gamma}$ at the LHC (see Fig. \ref{fig_2}). The luminosity spectrum is steeper at the LHeC and hence it is even larger at lower $W$ than at the LHC -- this is thanks to the  smaller electron mass, resulting in a significantly larger electron photon flux (at relative energies). One should note that the luminosity spectrum for the proton dissociative case is significantly flatter, and completely dominates at the LHeC the $\gamma\gamma$ collisions close to 1~TeV \cite{yy-paper}.

What is equally important, apart from the high $S_{\gamma\gamma}$ at the LHeC, are  favorable experimental conditions thanks to a very low event pileup \cite{lhec}. This will allow to apply powerful exclusivity selections using the calorimeters, which will cover very wide rapidity ranges and be sensitive to the neutral particles too, apart from a strong additional background suppression that will also allow to reconstruct  more complicated (than dilepton) final states. Finally, the photon-photon collisions at very large $W$ suffer at the LHC from a (very uncertain) suppression due to rescattering effects. That problem is completely absent at the LHeC as the incident electrons do not interact strongly.

A wide range of $\gamma\gamma$ processes will be studied at the LHeC with high significance, as: $\gamma\gamma\rightarrow\gamma\gamma$, $\gamma\gamma\rightarrow\tau\tau$,  $\gamma\gamma\rightarrow ZZ$, $\gamma\gamma\rightarrow WW$ and the anomalous $\gamma\gamma\rightarrow Z$, for example \cite{yy-paper}. It will be possible not only to go beyond the simplest (charged) di-lepton final states, but also to measure smaller signals thanks to both a more efficient signal detection and much more powerful background suppression.
\section{Conclusions}
The future Large Hadron Electron Collider offers fascinating perspectives for unique studies of high energy photon-photon interactions. It calls for comprehensive surveys to evaluate the full scientific potential of the $\gamma\gamma$  physics at the LHeC, as well as for initiating extensive studies of the relevant detector effects and requirements.

\end{document}